\documentclass[conference]{IEEEtran}
\IEEEoverridecommandlockouts
\usepackage{cite}
\usepackage{amsmath,amssymb,amsfonts}
\usepackage{algorithmic}
\usepackage{graphicx}
\usepackage{textcomp}
\usepackage{xcolor}

\usepackage{color}
\usepackage{bm}
\usepackage{amsthm}

\def\minwrt[#1]{\underset{#1}{\mathrm{minimize }}}
\def\argminwrt[#1]{\underset{#1}{\text{arg min }}}
\def\argmaxwrt[#1]{\underset{#1}{\text{arg max }}}
\def\maxwrt[#1]{\underset{#1}{\text{maximize }}}
\def\maxemphwrt[#1]{\underset{#1}{\text{\emph{maximize} }}}
\def\mminwrt[#1]{\underset{#1}{\mathrm{min }}}

\newcommand{\Tau}{\mathcal{T}}
\newcommand{\onevec}{\mathbf{1}}
\newcommand{\costfunc}{c}

\newtheorem{proposition}{Proposition}

\newcommand{\norm}[1]{\left\lVert#1\right\rVert}
\newcommand{\abs}[1]{\left|#1\right|}
\newcommand{\trace}[1]{\text{tr}\left(#1\right)}

\def\RR{{\mathbb{R}}}

\newcommand{\todor}[1]{\textcolor{red}{#1}} 

\newcommand{\omtplanfull}{\mathbf{M}}
\newcommand{\trashmass}{\mathbf{m}}
\newcommand{\trashcost}{\mathbf{c}}

\newcommand{\costmatfull}{\mathbf{C}}
\newcommand{\indvec}{\mathbf{q}}

\newcommand{\tdoaspertarget}{\tilde{R}}

\newcommand{\noise}{n}

\newcommand{\blambda}{\boldsymbol{\lambda}}
\newcommand{\bmu}{\boldsymbol{\mu}}

\newcommand{\bPhi}{\boldsymbol{\Phi}}
\newcommand{\bK}{\mathbf{K}}
\newcommand{\bk}{\mathbf{k}}

\newcommand{\bu}{\mathbf{u}}
\newcommand{\bv}{\mathbf{v}}
\newcommand{\bP}{\mathbf{P}}

\newcommand{\bx}{\mathbf{x}}
\newcommand{\by}{\mathbf{y}}

\begin{document}

\title{Multi-Source Localization and Data Association for Time-Difference of Arrival Measurements
}

\author{\IEEEauthorblockN{Gabrielle Flood}
\IEEEauthorblockA{\textit{Centre for Mathematical Sciences} \\
\textit{Lund University}, Lund, Sweden \\
firstname.lastname@math.lth.se}
\and
\IEEEauthorblockN{Filip Elvander}
\IEEEauthorblockA{\textit{Dept.\ of Information and Communications Engineering} \\
\textit{Aalto University}, Espoo, Finland \\
firstname.lastname@aalto.fi}
}

\maketitle

\begin{abstract}
In this work, we consider the problem of localizing multiple signal sources based on time-difference of arrival (TDOA) measurements. In the blind setting, in which the source signals are not known, the localization task is challenging due to the data association problem. That is, it is not known which of the TDOA measurements correspond to the same source.
%
%
Herein, we propose to perform joint localization and data association by means of an optimal transport formulation. The method operates by finding optimal groupings of TDOA measurements and associating these with candidate source locations. To allow for computationally feasible localization in three-dimensional space, an efficient set of candidate locations is constructed using a minimal multilateration solver based on minimal sets of receiver pairs. In numerical simulations, we demonstrate that the proposed method is robust both to measurement noise and TDOA detection errors. Furthermore, it is shown that the data association provided by the proposed method allows for statistically efficient estimates of the source locations. 
\end{abstract}



\section{Introduction}
%
Localization of signal sources occurs as a task in many signal processing applications, ranging from radar to audio signal processing, and constitutes a fundamental component in systems for tracking \cite{PotamitisCT04_12,NiuBVD12_48}, target identification \cite{LiS07_24}, and noise reduction \cite{GannotVMGO17_25}. In multi-sensor settings with unknown broad-band source signals, the spatial position of a source is often determined based on time-difference of arrival (TDOA) measurements corresponding to receiver-receiver pairs, obtained using, e.g., cross-correlation methods \cite{knapp1976generalized}. However, in scenarios with more than one signal source, localization is complicated by the lack of data labels. That is, for a number of sets of TDOAs corresponding to different receiver-receiver pairs, it is not known which have been generated by the same signal source. The grouping of TDOAs into subsets, each corresponding to a single source, is referred to as the data association problem and is essential to multi-source localization \cite{DangMHZ22_30,VelascoPMA16_64}. However, the two problems of data association and source localization are in general coupled, since TDOA association is often determined by goodness-of-fit to a given source location.
%
\begin{figure}
    \centering
     \includegraphics[trim={0.9cm 0.9cm 0.9cm 0.5cm},clip,width=0.47\columnwidth]{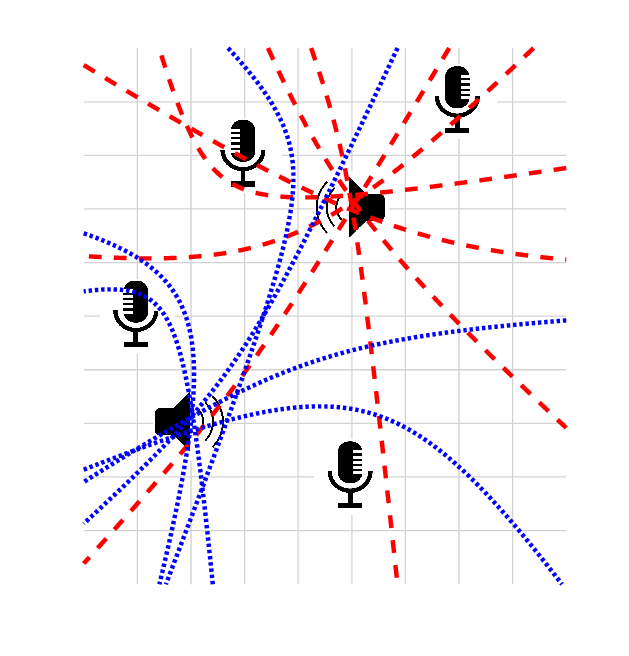}
     \includegraphics[trim={0.9cm 0.9cm 0.9cm 0.5cm},clip,width=0.47\columnwidth]{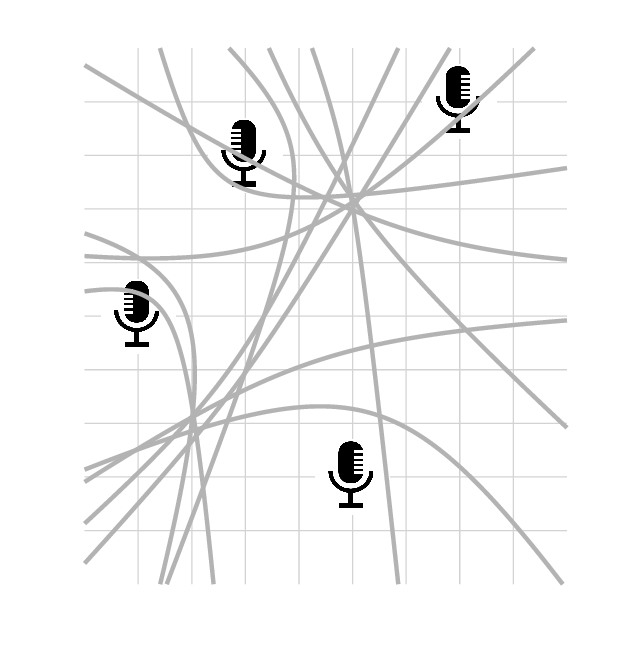}
     \vspace{-3mm}
     \caption{Two signal sources observed by four receivers, with hyperbolas corresponding to source locations consistent with TDOA measurements. Left: hyperbolas (TDOAs) labelled according to source. Right: unknown labels, corresponding to the localization and data association problem.}
     \label{fig:sslFramework}
\end{figure}
%
In the case when the target sensors transmit signals that are well separated in time, the problem can be solved as the single-sensor case, as the association is given by the order of the TDOAs. To solve for the position of a single transmitter among several receivers of known positions using TDOAs is called multilateration \cite{widdison2024review}, and the minimal configurations in 3D require four microphones. The position can also be found using more measurements, which might lead to a more accurate position, but with an increased risk of including corrupt data. It is common to use a setup where all measurements have the same reference receiver, \cite{chan1994simple}, however, there are also solvers without this requirement \cite{astrom2021extension}.

Previous works addressing multi-source localization in the absence of data labels include grid-based approaches originating from the compressed sensing paradigm \cite{JamaliL13_61, SchmitzMD15_cosera} (see also \cite{DongZZLWL23_11} for a similar approach using direction of arrival), association estimation based on enumerating all possible measurement combinations \cite{DangZC18_fusion}, as well as recent deep learning (DL) proposals \cite{AyubCZB23_27, SaadJZ22_192}. However, whereas the DL techniques suffer from being specific to scenarios consistent with their training data, the classical approaches are limited to low-dimensional settings, such as localization in the plane, due to the complexity incurred by gridding the space fine enough for reasonable retrieval robustness \cite{DangZ21_149,SundarSSC19_26}. Furthermore, even if location estimates are produced, the data association problem is not necessarily solved, preventing exploiting all available information for statistically efficient estimates \cite{JamaliL13_61}.
%

%
In this paper, we propose to jointly perform source localization and data association. In particular, we propose to phrase the association as an optimal transport (OT) problem \cite{villani08}, clustering TDOA measurements together by assigning them to a set of candidate locations.\footnote{Code will be released upon publication.} 
OT has previously successfully been applied to clustering problems in multi-static localization \cite{ElvanderKW21_eusipco}, as well as problems in spectral estimation \cite{ElvanderAKJ17_icassp, ElvanderJ20_68, Elvander23_icassp}.
In contrast to previous approaches, the candidate set used in the herein proposed work does not correspond to a gridding of the ambient space, but is formed using a multilateration scheme, which allows for localization also in three dimensions. Furthermore, for the OT problem we propose a novel combination of entropy (see \cite{Cuturi2013}) and sparsity regularization that allows us to derive an efficient and elegant solution algorithm. We demonstrate the robustness of the proposed method in the face of increasing sensor noise, as well as to missing and false TDOA measurements. The quality of the obtained data assocation is shown to be sufficient for statistically efficient source location estimates. 
\section{Source Localization}
Consider a scenario in which $S \in \mathbb{N}$ sources at unknown locations $s_j \in \RR^3$, for $j = 1,\ldots,S$, emit unknown signal waveforms that impinge on $R \in \mathbb{N}$ receivers at known locations $r_k \in \RR^3$, for $k = 1,\ldots,R$. The relative source-reciever distances are then enocded in TDOAs corresponding to a source-receiver-reciver triplet, i.e.,
\begin{align}
    \tau_{j,(k,\ell)} = \frac{\| s_j - r_k \| - \| s_j - r_\ell \|}{ \rho},
    \label{eq:tdoa_matrix_measurement}
\end{align}
where $\|\cdot\|$ denotes the Euclidean norm and $\rho$ is the speed of propagaition. We will, without loss of generality, further on assume that $\rho=1$. If the source signals are broadband and mutually uncorrelated, the TDOA $\tau_{j,(k,\ell)}$ may be estimated based on the cross-correlation between the receiver signals \cite{knapp1976generalized}. Herein, we assume that such estimated TDOAs are given, and they are modeled as random variables $\hat{\tau}_{j,(k,\ell)}$
\begin{align}\label{eq:tdoa_measurement}
    \hat{\tau}_{j,(k,\ell)} = \tau_{j,(k,\ell)} + \noise_{j,(k,\ell)},
\end{align}
where $\noise_{j,(k,\ell)}$ is an additive perturbation. For simplicity of the numerical experiements, but without loss of generality for the proposed method, we will assume $\noise_{j,(k,\ell)} \sim \mathcal{N}(0,\sigma^2)$, where the variance $\sigma^2$ will depend on the signal to noise ratio and the spectral characteristics of the source signals. With this model, the source positions $s_j$ can be identified if the number of receivers is $R\geq 4$.
However, for the case $S>1$, where the different signals are hard to distinguish in time, the estimation of $s_j$ is complicated by the fact that the source labels for the TDOA estimates are unknown. Hence, for two TDOA estimates $\hat{\tau}_{j,(k,\ell)}$ and $\hat{\tau}_{j',(k',\ell')}$ it is not known whether $j = j'$, i.e., if the two TDOAs correspond to the same source. Identifying groups of measurements, in this case TDOAs, corresponding to the same source is referred to as the data association problem, and if solved, the source locations $s_j$ can be estimated using, e.g., maximum likelihood methods.
%
%
%
%

However, in a blind estimation scenario in which the source signals are unknown, localization and data assocation become intrinsicly coupled problems. A 2D scenario is illustrated in Figure~\ref{fig:sslFramework}, showing two sources that are observed by four receivers. Here, the different hyperbolas correspond to the set of source locations consistent with a given receiver pair and TDOA. Note that due to the TDOA estimation errors, there are no source locations consistent with all measurements, i.e., the hyperbolas do not all intersect in a single point. The joint data association and source localization problem is shown in the right-most plot where both source locations and data labels (associations) are missing.
\section{Method Outline}
Herein, we propose to jointly solve the data association and localization problems using an optimal transport (OT) formulation. In particular, the solution to the OT problem will provide an optimal association between the set of available TDOA estimates and a set of candidate source locations. This set of candidate locations is constructed be means of a multilateration solver, avoiding the need for extensive gridding of 3D space. Our proposed strategy is outlined as follows:
%
%
%
%
\begin{enumerate} 
    \item Pick a set of receiver pairs suitable to the chosen multilateration solver. For each pair, create all combinations of TDOAs, with one from each pair. This will include erroneous combinations, as well as the correct ones.
    \item Solve the multilateration problem for each TDOA combination. Collect all different solutions in a set $\Omega_i$ of candidate source positions $x \in \RR^3$.
    \item Repeat for more sets of pairs to build the full candidate set $\Omega = \cup_{i = 1}^K \Omega_i$, where $K\geq 1$ is a design parameter. 
    \item Solve the OT problem to find the association between TDOAs $\Tau = \{ \hat{\tau}_{j,(k,\ell)} \}_{j,k,\ell}$ and candidate source positions in $\Omega$. This also finds the $S$ best candidate positions.
    \item Given the estimated data association as well as identified source positions, compute refined location estimates using a local search. Before, the association can be re-estimated using only the top candidates from step 4. 
\end{enumerate}
The candidate set construction is presented in Section~\ref{sec:multilateration}, whereas the OT problem for data assocation is presented in Section~\ref{sec:omt}, together with an efficient solution algorithm. The proposed method is reminiscent of our earlier work \cite{ElvanderKW21_eusipco} for multi-static time of arrival based localization, however, the multilateration strategy proposed here, together with the novel OT solver, allows for addressing problems in 3D.
%

\section{Candidate Set Construction}
\label{sec:multilateration}

The candidate set is constructed by solving a number of multilateration problems. To avoid high dependencies on and sensitivity to noise in a single receiver we choose a multilateration formulation without a fixed reference receiver. Hence, we use Equation \eqref{eq:tdoa_matrix_measurement}, and allow $k$ and $\ell$ to be different for each measurement.
In this case, it is enough to use three measurement equations, given that these are independent (the set of indices $\{ k_1,\ell_1,k_2,\ell_2,k_3,\ell_3 \}$ must contain at least four unique indices). We choose to use the minimal multiateration solver presented in \cite{astrom2021extension}. However, the suggested method will work for other solvers as well, but the number of TDOA measurements might vary. Therefore, we denote this value by $P$, even if $P=3$ will be used throughout the paper.

Thus, in order to construct a candidate set $\Omega_i$ of source positions, first pick a minimal (or suitable) set of $P$ receiver pairs, corresponding to $P$ sets of TDOAs. Let these sets be denoted 
$\Psi_{k,\ell} =  \{ \hat{\tau}_{j,(k,\ell)} \}_{j}$. Thereafter, the set of all combinations of TDOAs, $\Lambda_i = \Psi_{k_1,\ell_1} \times \hdots \times \Psi_{k_P,\ell_P}$, is created.
%
Each element in $\Lambda_i$ is a $P$-tuple of TDOAs, and yields -- using the multilateration method -- one or more candidate source positions. The set of all such candidate positions for receiver pair set number $i$ is collected in the set $\Omega_i$,
%
but in some cases, some of the solutions given by the solver can be discarded. For the solver in \cite{astrom2021extension} we discard any solution that does not match the TDOAs used, and also solutions for which the imaginary part is too large.

Furthermore, several sets of receiver pairs can be considered, as to form $\Omega_i$, $i = 1,\ldots,K$. Ideally, these sets should be disjoint as to be robust to missing data (missed detections), but may also be chosen randomly. The full set of candidate positions is constructed as $\Omega = \cup_{i=1}^K \Omega_i$, whereafter duplicates are removed. Further on, the elements in $\Omega$ will be referred to as $x_j$, as the order of the candidates is not important. 

\section{Association Problem}
\label{sec:omt}
Given the set $\Omega$ of candidate locations, we propose to both localize the sources $s_j$ and identify its corresponding TDOAs by finding an optimal association between the sets $\Omega$ and $\Tau$. In particular, reminiscent of the approach in \cite{ElvanderKW21_eusipco}, we propose to do this by means of an OT problem. Building on the discrete Monge-Kantorovich OT problem \cite{villani08}, we would ideally solve
\begin{equation} \label{eq:ip_ot}
\begin{aligned}
    \minwrt[\substack{\omtplanfull \in \{0,1 \}^{ |\Tau| \times |\Omega|},\: \indvec \in \{0,1\}^{|\Omega|}}]\quad& \langle \costmatfull,\omtplanfull\rangle = \trace{\costmatfull^T\omtplanfull } \\
    \text{s.t. }\quad& \omtplanfull \onevec_{|\Omega|} = \onevec_{|\Tau|} \\
    &\omtplanfull^T\onevec_{|\Tau|} = \tdoaspertarget \onevec_{|\Omega|} \\
    &\indvec^T \onevec_{|\Omega|} = S ,\;\; \omtplanfull \leq \onevec_{|\Tau|} \indvec^T,
\end{aligned}
\end{equation}
where the last constraint is to be interpreted elementwise, and where $\tdoaspertarget = R(R-1)/2$. Here, $\omtplanfull$ is the transport or association plan describing the association between TDOAs and candidate sources, s.t., $[\omtplanfull]_{i,j} = 1$ if TDOA $\hat{\tau}_i$ is assocated with source candidate $x_j$, and zero otherwise. The quality of the association is measured by the objective function, with $\costmatfull \in \RR^{|\Tau| \times |\Omega|}$ constructed as $[\costmatfull]_{i,j} = c(x_j,\hat{\tau}_i;(k,\ell))$
%
%
where $\costfunc: \RR^3 \times \RR \rightarrow \RR_+$ is a consistency measure, or ground cost, for a candidate source position and a TDOA corresponding to a receiver pair. In this work, we will use
\begin{align} \label{eq:individual_cost}
    c(x,\tau;(k,\ell))= \left( \norm{x- r_k }_2 - \norm{x-r_\ell}_2 - \tau \right)^2,
\end{align}
which corresponds to the assumption that TDOA estimates are Gaussian random variables.
Furthermore, the first and second constraints in \eqref{eq:ip_ot} ensure that each TDOA is associated with a target, and that each target is associated with $\tdoaspertarget$ TDOAs (the number of receiver pairs), respectively. 
Here, $\onevec_{m}$ denotes a $m\times 1$ vector of all 1's.
The last two constraints ensure that only $S$ candidate sources are selected. The $S$-sparse vector $\indvec$ indicates which candidate sources are selected.

As may be noted, the minimal problem in \eqref{eq:ip_ot} is combinatorial due to the integer constraints. Furthermore, it may be noted that \eqref{eq:ip_ot} assumes that there are neither missing nor false  TDOA measurements. In order to address this, and to arrive at a computationally efficient method for solving the association problem, we propose the following relaxation:
\begin{equation} \label{eq:entropy_ot}
\begin{aligned}
    \minwrt[\substack{\omtplanfull \in \RR^{ |\Tau| \times |\Omega|} \\ \trashmass \in \RR^{|\Tau|}}]\: & \langle \costmatfull,\omtplanfull\rangle \!+\! \langle \trashcost,\trashmass\rangle \!+\! \epsilon (D(\omtplanfull)\!+\!D(\trashmass)) \!+\!\eta \norm{\omtplanfull}_{\infty,1} \\
    \text{s.t. }\: & \omtplanfull \onevec_{|\Omega|} +\trashmass = \onevec_{|\Tau|} ,\;\;\omtplanfull^T\onevec_{|\Tau|} \leq \tdoaspertarget \onevec_{|\Omega|},
\end{aligned}
\end{equation}
where $D(\omtplanfull) = \sum_{i,j}[\omtplanfull]_{i,j} \log [\omtplanfull]_{i,j} - [\omtplanfull]_{i,j}+1 $ is an entropic regularization term\footnote{Entropy regularization, applied to the standard OT problem, was first introduced in \cite{Cuturi2013}.} (with $D(\trashmass)$ defined analogously), $\eta$ and $\epsilon$ are positive parameters, and where the mixed norm is
\begin{align*}
    \norm{\omtplanfull}_{\infty,1} = \sum_{j} \max_{i} \abs{[\omtplanfull]_{i,j}}.
\end{align*}
In \eqref{eq:entropy_ot}, we have introduced an additional transportation vector $\trashmass \in \RR^{|\Tau|}$ that allows for so-called unbalanced OT \cite{peyre2019computational} by associating TDOAs with a void, or trash, source to a cost $\trashcost \in \RR^{|\Tau|}$. Also, changing equality to inequality in the second constraint allows for missing TDOAs. Furthermore, the integer constraints have been relaxed to allow the elements of $\omtplanfull$ and $\trashmass$ to take on any real value. However, it may be readily verified that any feasible point with non-infinite objective value satisfies $\omtplanfull \in [0,1]^{|\Tau| \times |\Omega|}$ and $\trashmass \in [0,1]^{|\Tau|}$. Lastly, the sparsity-promoting constraints in \eqref{eq:ip_ot} have here been replaced with a sparsity-promoting penalty\footnote{Note that the two last constraints of \eqref{eq:ip_ot} are equivalent to $\norm{\omtplanfull}_{\infty,1} = S$.}. Relating to \eqref{eq:ip_ot}, the source indicator vector $\indvec$ is constructed as the column-wise maxima of the optimal $\omtplanfull$. The appeal of the relaxed problem \eqref{eq:entropy_ot} is that it allows for an efficient solution algorithm via its convex dual.
The following proposition holds.
\begin{proposition}
The unique optimal $(\omtplanfull,\trashmass)$ is represented as\looseness=-1
\begin{align*}
    %
   %
   \omtplanfull = \bP \odot \bK \odot \left(\bv \otimes \bu\right) ,\;\; \trashmass = \bk \odot \bv
\end{align*}
where $\bK =\exp\left(-\frac{1}{\epsilon}\costmatfull \right)$, $\bk = \exp\left(-\frac{1}{\epsilon}\trashcost \right) $, and
\begin{align*}
    \bu = \exp\left( -\frac{1}{\epsilon} \bmu \right),\;\; \bv = \exp\left( \frac{1}{\epsilon} \blambda \right) ,\;\; \bP = \exp\left( \frac{1}{\epsilon} \bPhi \right),
\end{align*}
where $\blambda$, $\bmu$, and $\bPhi$ solve the dual problem
\begin{equation} \label{eq:ot_dual}
\begin{aligned}
    \minwrt[\blambda \in \RR^{|\Tau|},\,\bmu \in \RR_+^{|\Omega|},\, \bPhi \in \RR^{|\Tau| \times |\Omega|}] & \quad\epsilon\left\langle \bP\odot\bK,\bv \otimes \bu   \right\rangle + \epsilon \langle \bk, \bv \rangle  \\[-5pt]&- \langle \onevec_{|\Tau|}, \blambda \rangle +\tdoaspertarget\langle \onevec_{|\Omega|}, \bmu \rangle
    \\ \mathrm{s.t. } &\quad \norm{\bPhi}_{1,\infty} \leq \eta,
\end{aligned}
\end{equation}
where $\norm{\bPhi}_{1,\infty} = \max_{j} \sum_{i} \abs{[\bPhi]_{i,j}}$. Here, exponentiation is elementwise, and $\otimes$ and $\odot$ denote the Kronecker and Hadamard products, respectively. Furthermore, for any non-negative $\costmatfull$ and $\trashcost$, strong duality holds.
\end{proposition}
The proof is omitted due to the page limitation. However, it may be noted that Slater's condition holds for \eqref{eq:entropy_ot}, and as the objective is bounded from below for non-negative $\costmatfull$ and $\trashcost$, strong duality follows.  Furthermore, the dual problem \eqref{eq:ot_dual} can be solved efficiently using block-coordinate descent, which can be interpreted as generalized Sinkhorn iterations \cite{KarlssonR17_10,elvander2020multi}. In particular, in iteration $j$, the variables are updated as
\begin{align*}
    \blambda^{(j)} &= -\epsilon \log \left( \bP^{(j-1)} \odot \bK \bu^{(j-1)} \right) 
    \\ \bmu^{(j)} &= \!\epsilon \!\left( \!\log\!\left( \bP^{(j-1)^T} \!\odot \! \bK^T \bv^{(j)} \right) \!-\! \log\! \tilde{R} \onevec_{|\Omega|}\!\right)_+
    \\ \left[\bPhi^{(j)}\right]_{:,\ell} \!&=\! -\epsilon\,\Gamma\!\left( \left[\bK \odot \bv^{(j)} \otimes \bu^{(j)}\right]_{:,\ell} , \eta/\epsilon \right) ,\, \ell \!=\! 1,\ldots, |\Omega|
\end{align*}
where the $\log$-operations are elementwise, $(\cdot)_+$ is elementwise truncation at zero, and $\Gamma: \RR^{|\Tau|}\times \RR_+ \to \RR^{|\Tau|}$ is defined as
\begin{align}
    \Gamma(\by, p) &= \argminwrt[\bx] \langle \exp(\bx), \by \rangle \;,\; \text{s.t. } \norm{\bx}_1 \leq p.
\end{align}
Due to space constraints, we omit the proof of these updates. For computing $\Gamma$, we use the solver from \cite{haaslerE24_css_arxiv}.
It may be verified that all computations can be performed in log-domain using log-sum-exp stabilization (see, e.g., \cite{peyre2019computational}). Thus, the entropy regularization parameter $\epsilon$ can be selected arbitrarily small without causing numerical overflow in the iterations, and may be regarded as a parameter for the solution algorithm rather than for the primal problem \eqref{eq:entropy_ot} itself. In our numerical experiments, we set $\epsilon = 10^{-7}$ to closely approximate the corresponding OT problem without entropy regularization.

The overall computational complexity of the problem is $\mathcal{O}\left( |\Omega||\Tau| \log |\Tau| \right)$, resulting from the updates of $\bPhi$. Here, the complexity for evaluating $\Gamma$ is $\mathcal{O}\left( |\Tau| \log |\Tau| \right)$ \cite{haaslerE24_css_arxiv}.
%
Since the optimal $\omtplanfull$ gives an estimate of the association between TDOAs and candidate sources, the location estimates can be refined by a local non-linear search implementing, e.g., a maximum likelihood estimator exploiting all the available data. This will be demonstrated in the numerical section.
%
%
%
\section{Experimental Validation}
In this section, we evaluate the performance of the proposed method, and in particular its robustness to perturbations in the form of TDOA measurement noise as well as to false and missing TDOAs.
%
%
For all experiments, $R=12$ receiver and $S=3$ source positions were in each simulation generated randomly in a $10 \times 10 \times 2$~m room. The implied TDOAs were computed for each receiver pair, whereafter these were perturbed with Gaussian noise $\noise \sim \mathcal{N}(0,\sigma^2)$, where $\sigma$ is the standard devation.  The number of minimal sets of receiver pairs was set to $K=3$ and the two index sets were chosen randomly, but we ensured that no receiver pair was used twice. 
The solver from \cite{astrom2021extension} was used for multilateration according to Section \ref{sec:multilateration} and duplicates whose distance to another candidate were smaller than 1 cm were exluded from the full candidate set $\Omega = \cup_{i=1}^3 \Omega_i$. 
For the association problem, we used the ground cost in \eqref{eq:individual_cost}, with the cost $\trashcost$ for the void source being set to the 95th percentile of the empirical distribution of the costs corresponding to the candidates.
With $\eta = 1$, the estimated source positions were determined as the $S$ dominant columns of $\omtplanfull$ solving \eqref{eq:entropy_ot}. Using the association provided by $\omtplanfull$, the location estimates were then refined using a local search minimizing \eqref{eq:individual_cost} for each source individually, using Matlab's \textit{fminsearch}. 
To increase the chances for identifiable source positions, in a statistical sense, for the multilateration solver, we re-estimated the sources using new index sets if the root Cram\'er-Rao lower bound (CRLB) was higher than 20 m for all $K$ index sets with found source positions.
As a performance measure, we present the average distance from the estimate to ground truth source (in meters), as well as the association rate, i.e., the fraction of TDOAs that were correctly associated with either its corresponding source or the void source.

As reference, we include estimates by the SMTL method \cite{JamaliL13_61} using the setting recommended by the authors. As it is not computationally feasible to run this method with a uniform gridding of the room, we use our candidate set as grid. Also, we include the CRLB, corresponding to the setup with all the available TDOA measurements, as an absolute benchmark.
%
%
%

%
\begin{figure}
    \centering
     \includegraphics[trim={0cm 0cm 0cm 0cm},clip,width=\columnwidth]{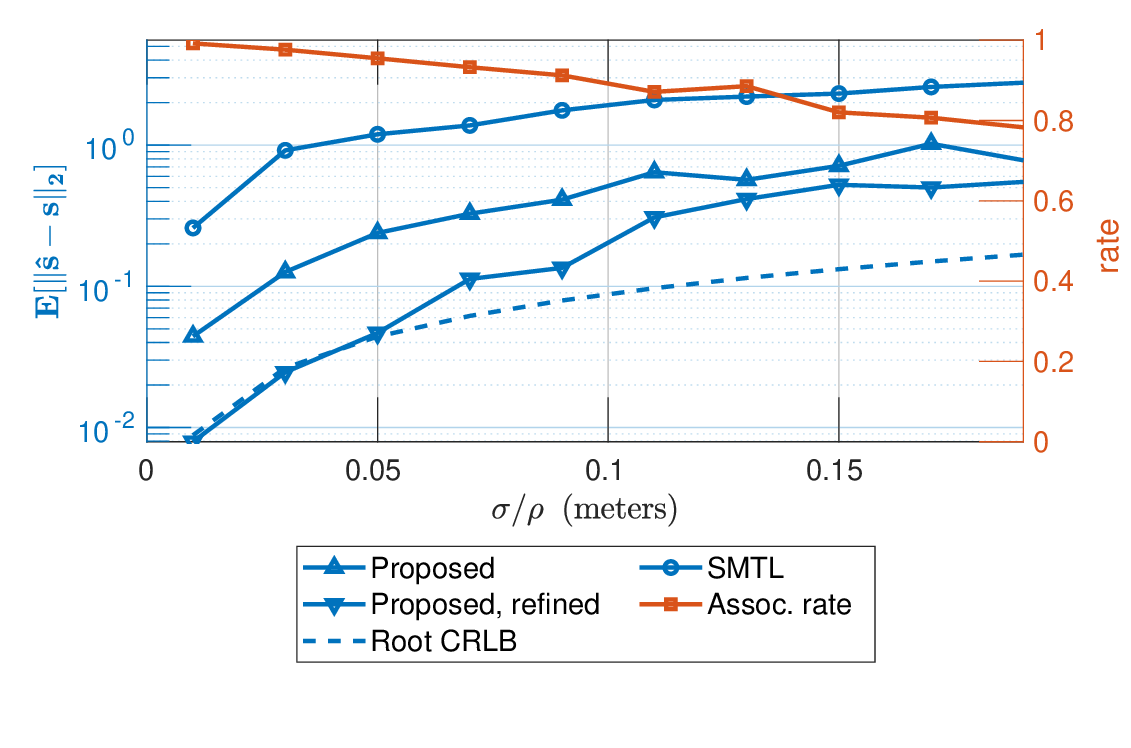}
     \vspace{-10mm}\caption{Expected localization error (left axis) and the rate of correctly associated TDOA measurements (right axis) for varying noise level $\sigma$.}
     \label{fig:err_wrt_pertub}
\end{figure}
%
\subsection{Noise sensitivity}
Here, we evaluate the performance with respect to the size of the measurement error $\noise$ in \eqref{eq:tdoa_measurement}. The standard deviation $\sigma$ of the error was varied between $[0.01, 0.19]$, and there were no false or missing TDOA measurements. The average results over 100 simulations are shown in Figure~\ref{fig:err_wrt_pertub}, with the left and right y-axes showing the average Euclidean distance between estimates and ground truth sources, and the average association rate for the proposed method, respectively. As can be seen, the proposed method displays considerable robustness to the TDOA noise. Also, the number of TDOAs correctly associated with their respective sources is high enough as to achieve statistical efficiency in the refinement step. It may here be noted that the statistical accuracy of the estimates without refinement is limited due to the small number of measurements used in the multilateration step.

%
%

\subsection{Missing and False TDOA Measurements}
\begin{figure}
    \centering
     \includegraphics[trim={0cm 0cm 0cm 0cm},clip,width=\columnwidth]{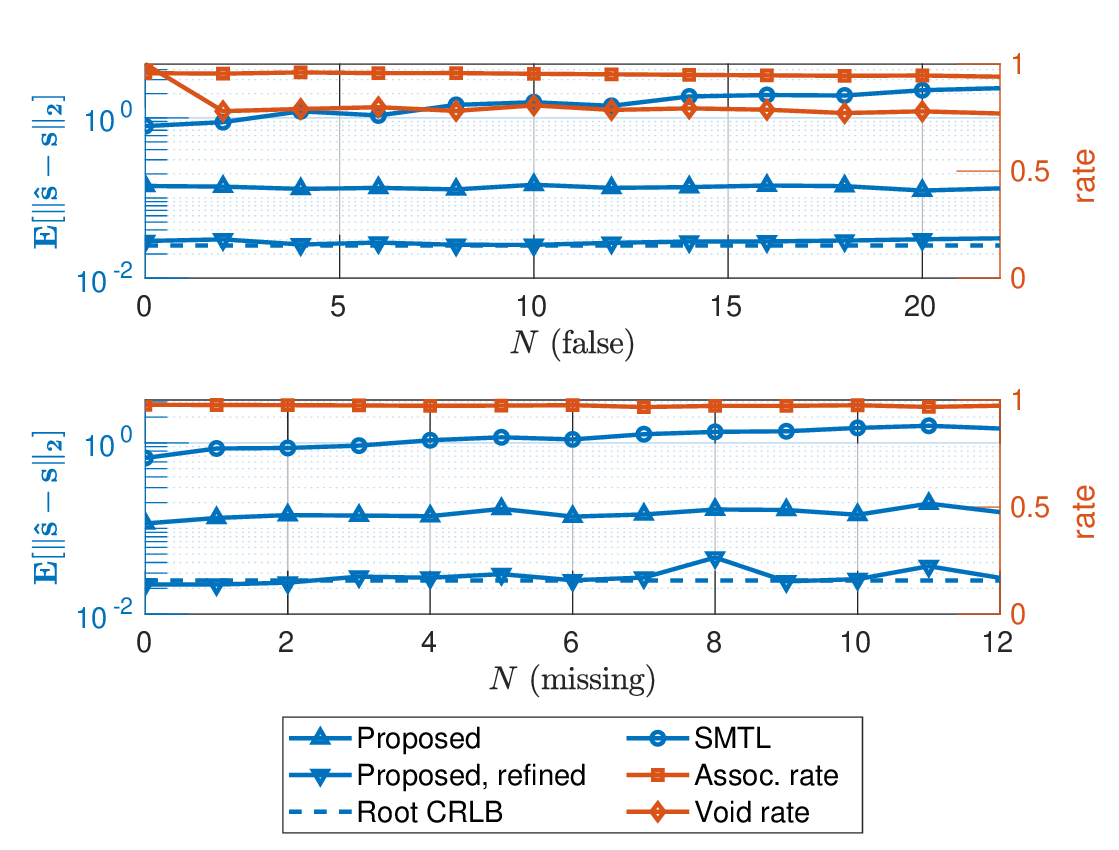}
\vspace{-5mm}     \caption{Expected localization error (left axis) and the rate of correctly associated TDOA measurements (right axis) with respect to false and missing measurements, on the top and bottom, respectively. For false measurements the association rate for false-to-void source is included.}
     \label{fig:err_wrt_false_missing}
\end{figure}

When evaluating the robustness to false and missing TDOA measurements, we fixed the noise standard deviation at $\sigma = 0.03$. 
In the case of false TDOAs, we $N \in \mathbb{N}$ times chose a receiver pair at random and added a false TDOA, drawn uniformly in the range of the corresponding pairs already existing TDOAs, yielding in total $N$ false measurements.
For the case of missing TDOAs, we randomly deleted measurements using an analogue procedure. The maximum considered value $N = 22$ corresponds to approximately 10\% of the available TDOA measurements. The average results over 100 simulations are displayed in Figure~\ref{fig:err_wrt_false_missing}, with the upper and lower plots showing results for false and missing TDOA measurements, respectively. As may be noted, although false and missing values affects the association rate, the proposed method yields accurate estimates for moderately large $N$.
%

%
%

%
\section{Conclusions}
\vspace{-1mm}
In this work, we presented a method for joint multi-source localization and data association for TDOA measurements. First, several minimal multilateration problems are solved, and all solutions are considered candidate source positions. Thereafter, an optimal transport formulation, with entropy and sparsity regularization, is used to associate the TDOAs with the desired amount of candidates while also finding the best source positions. The obtained association then allows for a refinement stage achieving statistical effciency. Experiments show that the proposed method is robust to measurement noise, as well as false or missing TDOA detections. 
%
\vspace{-1mm}
\bibliographystyle{IEEEtran}
\bibliography{IEEEabrv,refs}

\begin{thebibliography}{10}
\providecommand{\url}[1]{#1}
\csname url@samestyle\endcsname
\providecommand{\newblock}{\relax}
\providecommand{\bibinfo}[2]{#2}
\providecommand{\BIBentrySTDinterwordspacing}{\spaceskip=0pt\relax}
\providecommand{\BIBentryALTinterwordstretchfactor}{4}
\providecommand{\BIBentryALTinterwordspacing}{\spaceskip=\fontdimen2\font plus
\BIBentryALTinterwordstretchfactor\fontdimen3\font minus
  \fontdimen4\font\relax}
\providecommand{\BIBforeignlanguage}[2]{{%
\expandafter\ifx\csname l@#1\endcsname\relax
\typeout{** WARNING: IEEEtran.bst: No hyphenation pattern has been}%
\typeout{** loaded for the language `#1'. Using the pattern for}%
\typeout{** the default language instead.}%
\else
\language=\csname l@#1\endcsname
\fi
#2}}
\providecommand{\BIBdecl}{\relax}
\BIBdecl

\bibitem{PotamitisCT04_12}
I.~Potamitis, H.~Chen, and G.~Tremoulis, ``{T}racking of multiple moving
  speakers with multiple microphone arrays,'' \emph{IEEE Trans. Speech Audio
  Process.}, vol.~12, no.~5, pp. 520--529, Sept 2004.

\bibitem{NiuBVD12_48}
R.~Niu, R.~S. Blum, R.~K. Varshney, and A.~L. Drozd, ``Target {L}ocalization
  and {T}racking in {N}oncoherent {M}ultiple-{I}nput {M}ultiple-{O}utput
  {R}adar {S}ystems,'' \emph{IEEE Trans. Aerosp. Electron. Syst.}, vol.~48,
  no.~2, pp. 1466--1489, 2012.

\bibitem{LiS07_24}
J.~Li and P.~Stoica, ``{MIMO} {R}adar with {C}olocated {A}ntennas,'' \emph{IEEE
  Signal Process. Mag.}, vol.~24, no.~5, pp. 106--114, 2007.

\bibitem{GannotVMGO17_25}
S.~Gannot, E.~Vincent, S.~Markovich-Golan, and A.~Ozerov, ``A {C}onsolidated
  {P}erspective on {M}ultimicrophone {S}peech {E}nhancement and {S}ource
  {S}eparation,'' \emph{IEEE/ACM Trans. Audio, Speech, Lang. Process.},
  vol.~25, no.~4, pp. 692--730, 2017.

\bibitem{knapp1976generalized}
C.~Knapp and G.~Carter, ``The generalized correlation method for estimation of
  time delay,'' \emph{IEEE Trans. Acoust., Speech, Signal Process.}, vol.~24,
  no.~4, pp. 320--327, 1976.

\bibitem{DangMHZ22_30}
X.~Dang, W.~Ma, E.~A.~P. Habets, and H.~Zhu, ``Tdoa-based robust sound source
  localization with sparse regularization in wireless acoustic sensor
  networks,'' \emph{IEEE/ACM Trans. Audio, Speech, Lang. Process.}, vol.~30,
  pp. 1108--1123, 2022.

\bibitem{VelascoPMA16_64}
J.~Velasco, D.~Pizarro, J.~Macias-Guarasa, and A.~Asaei, ``Tdoa matrices:
  Algebraic properties and their application to robust denoising with missing
  data,'' \emph{IEEE Trans. Signal Process.}, vol.~64, no.~20, pp. 5242--5254,
  2016.

\bibitem{widdison2024review}
E.~Widdison and D.~G. Long, ``A review of linear multilateration techniques and
  applications,'' \emph{IEEE Access}, 2024.

\bibitem{chan1994simple}
Y.~T. Chan and K.~C. Ho, ``A simple and efficient estimator for hyperbolic
  location,'' \emph{IEEE transactions on signal processing}, vol.~42, no.~8,
  pp. 1905--1915, 1994.

\bibitem{astrom2021extension}
K.~Åström, M.~Larsson, G.~Flood, and M.~Oskarsson, ``Extension of
  time-difference-of-arrival self calibration solutions using robust
  multilateration,'' in \emph{European Signal Process. Conf.}, 2021, pp.
  870--874.

\bibitem{JamaliL13_61}
H.~Jamali-Rad and G.~Leus, ``Sparsity-aware multi-source tdoa localization,''
  \emph{IEEE Trans. Signal Process.}, vol.~61, no.~19, pp. 4874--4887, 2013.

\bibitem{SchmitzMD15_cosera}
J.~Schmitz, R.~Mathar, and D.~Dorsch, ``Compressed time difference of arrival
  based emitter localization,'' in \emph{2015 3rd International Workshop on
  Compressed Sensing Theory and its Applications to Radar, Sonar and Remote
  Sensing (CoSeRa)}, 2015, pp. 263--267.

\bibitem{DongZZLWL23_11}
N.~Dong, L.~Zhang, H.~Zhou, X.~Li, S.~Wu, and X.~Liu, ``Two-stage fast matching
  pursuit algorithm for multi-target localization,'' \emph{IEEE Access},
  vol.~11, pp. 66\,318--66\,326, 2023.

\bibitem{DangZC18_fusion}
X.~Dang, H.~Zhu, and Q.~Cheng, ``Multiple sound source localization based on a
  multi-dimensional assignment model,'' in \emph{Int. Conf. Inf. Fusion}, 2018,
  pp. 1732--1737.

\bibitem{AyubCZB23_27}
M.~S. Ayub, J.~Chen, A.~Zaman, and J.~Bai, ``Deep attention aware feature
  learning for data association in multiple source localization,'' \emph{IEEE
  Communications Letters}, vol.~27, no.~1, pp. 125--129, 2023.

\bibitem{SaadJZ22_192}
M.~S. Ayub, C.~Jianfeng, and A.~Zaman, ``Multiple acoustic source localization
  using deep data association,'' \emph{Appl. Acoust.}, vol. 192, p. 108731,
  2022.

\bibitem{DangZ21_149}
X.~Dang and H.~Zhu, ``{A feature-based data association method for multiple
  acoustic source localization in a distributed microphone array},'' \emph{J.
  Acoust. Soc. Am.}, vol. 149, no.~1, pp. 612--628, 01 2021.

\bibitem{SundarSSC19_26}
H.~Sundar, T.~V. Sreenivas, and C.~S. Seelamantula, ``{TDOA}-based multiple
  acoustic source localization without association ambiguity,'' \emph{IEEE/ACM
  Trans. Audio, Speech, Lang. Process.}, vol.~26, no.~11, pp. 1976--1990, 2018.

\bibitem{villani08}
C.~Villani, \emph{Optimal Transport: Old and New}.\hskip 1em plus 0.5em minus
  0.4em\relax Berlin: Springer, 2008.

\bibitem{ElvanderKW21_eusipco}
F.~Elvander, J.~Karlsson, and T.~van Waterschoot, ``Convex {C}lustering for
  {M}ultistatic {A}ctive {S}ensing via {O}ptimal {M}ass {T}ransport,'' in
  \emph{European Signal Process. Conf.}, 2021.

\bibitem{ElvanderAKJ17_icassp}
F.~Elvander, S.~I. Adalbj\"ornsson, J.~Karlsson, and A.~Jakobsson, ``{U}sing
  {O}ptimal {T}ransport for {E}stimating {I}nharmonic {P}itch {S}ignals,'' in
  \emph{IEEE Int. Conf. Acoust. Speech, Signal Process.}, 2017, pp. 331--335.

\bibitem{ElvanderJ20_68}
F.~Elvander and A.~Jakobsson, ``Defining {F}undamental {F}requency for {A}lmost
  {H}armonic {S}ignals,'' \emph{IEEE Trans. Signal Process.}, vol.~68, pp.
  6453--6466, 2020.

\bibitem{Elvander23_icassp}
F.~Elvander, ``Estimating {I}nharmonic {S}ignals with {O}ptimal {T}ransport
  {P}riors,'' in \emph{IEEE Int. Conf. Acoust., Speech, Signal Process.}, 2023.

\bibitem{Cuturi2013}
M.~Cuturi, ``Sinkhorn distances: Lightspeed computation of optimal transport,''
  in \emph{Adv Neural Inf Process Syst}, 2013, pp. 2292--2300.

\bibitem{peyre2019computational}
G.~Peyr{\'e}, M.~Cuturi \emph{et~al.}, ``Computational optimal transport: With
  applications to data science,'' \emph{Found. Trends Mach. Learn.}, vol.~11,
  no. 5-6, pp. 355--607, 2019.

\bibitem{KarlssonR17_10}
J.~Karlsson and A.~Ringh, ``Generalized {S}inkhorn iterations for regularizing
  inverse problems using optimal mass transport,'' \emph{SIAM J. Imag. Sci.},
  vol.~10, no.~4, pp. 1935--1962, 2017.

\bibitem{elvander2020multi}
F.~Elvander, I.~Haasler, A.~Jakobsson, and J.~Karlsson, ``Multi-marginal
  optimal transport using partial information with applications in robust
  localization and sensor fusion,'' \emph{Signal Processing}, vol. 171, p.
  107474, 2020.

\bibitem{haaslerE24_css_arxiv}
I.~Haasler and F.~Elvander, ``Multi-frequency tracking via group-sparse optimal
  transport,'' \emph{arXiv preprint arXiv:2402.19345}, 2024.

\end{thebibliography}
\end{document}